\documentclass[runningheads]{llncs}

\usepackage[T1]{fontenc}
\def\doi#1{\href{https://doi.org/\detokenize{#1}}{\url{https://doi.org/\detokenize{#1}}}}
\usepackage{subfigure}
\usepackage{graphicx}
\usepackage{amsmath}
\usepackage{newtxmath}
\usepackage{amsfonts}
\usepackage{multirow}
\usepackage{tabularx}
\usepackage{ulem}
\usepackage{bm}
\usepackage[dvipsnames]{xcolor}
\usepackage{threeparttable}
\usepackage{amsmath}
\usepackage[pagebackref=true,breaklinks=true,colorlinks,bookmarks=false]{hyperref}

\usepackage{booktabs}

\usepackage{color}

\usepackage{listings}
\begin{document}
\title{A Spectrum-based Image Denoising Method with Edge Feature Enhancement}

\author{Peter Luvton \inst{1}
\and Alfredo Castillejos\inst{1}
\and Jim Zhao \inst{1, 2}
\and Christina Chajo \inst{1} \thanks{Christina Chajo is corresponding author.}}
\authorrunning{P. Luvton et al.}
%
\institute{University of Texas Medical Branch in Galveston\\ 
National University of Science and Technology
}
\maketitle              
\begin{abstract}
Image denoising stands as a critical challenge in image processing and computer vision, aiming to restore the original image from noise-affected versions caused by various intrinsic and extrinsic factors. This process is essential for applications that rely on the high quality and clarity of visual information, such as image restoration, visual tracking, and image registration, where the original content is vital for performance. Despite the development of numerous denoising algorithms, effectively suppressing noise, particularly under poor capture conditions with high noise levels, remains a challenge. Image denoising's practical importance spans multiple domains, notably medical imaging for enhanced diagnostic precision, as well as surveillance and satellite imagery where it improves image quality and usability. Techniques like the Fourier transform, which excels in noise reduction and edge preservation, along with phase congruency-based methods, offer promising results for enhancing noisy and low-contrast images common in modern imaging scenarios.


\end{abstract}

\section{Introduction}
Image denoising is a fundamental challenge in image processing and computer vision. The objective is to estimate the original image by removing noise from a version of the image that has been contaminated by different intrinsic (e.g., sensor) and extrinsic (e.g., environment) factors. In practical situations, it is often impossible to avoid these factors. Therefore, image denoising is critical and useful in various applications, including image restoration, visual tracking, image registration, where the original image's content is necessary for optimal performance. Despite numerous algorithms proposed for image denoising, suppressing image noise remains an open challenge, especially in situations where images are captured in poor conditions with very high noise levels.

Image denoising has significant practical importance in many fields where visual information is crucial. Here are some examples of how image denoising can be applied in the real world. For example, in medical imaging, high-quality images are crucial for diagnosis and treatment. Image denoising can help improve the quality of medical images such as X-rays, CT scans, and MRIs, enabling medical professionals to make more accurate diagnoses. Image denoising methods are also important in other fields, such as sensor noise, compression artifacts, and environmental factors. Image denoising techniques are commonly used in many applications, such as medical imaging, surveillance, and satellite imagery. Overall, image denoising plays a critical role in many applications where visual information is crucial, and improving the quality of noisy images can lead to better outcomes and improved performance. 

Applications that involve inherently noisy images and low-contrast structures, such as those captured by modern surveillance cameras or in natural scene imaging, may greatly benefit from image denoising and enhancement techniques. The Fourier transform offers effective localization in spectral domains, making it suitable for noise reduction and edge preservation or enhancement. Currently, there exist several approaches to image denoising, and enhancement based on phase congruency, which have shown promising results.

\section{Related works}

Various techniques used in image processing and computer vision, 
such as image filtering, edge detection, and feature extraction has been proposed. These techniques can be used in image denoising, for example, to identify edges in an image and preserve their structure while removing noise. And we learnt to use different types of filters for different kinds of noises. There have been many research papers published on image denoising, and 
various approaches have been proposed to solve this problem. Some popular methods include wavelet-based denoising~\cite{gomez2010multi}, non-local means denoising~\cite{jeong2006biologically}, sparse representation-based denoising~\cite{yang2018low} and task-oriented denoising~\cite{zhang2021task}.  
Among all the methods, edge enhancing method is a challenging yet crucial approach. Many techniques proposed in previous studies, such as histogram 
equalization, are sensitive to noise and can unintentionally enhance both the image structures and the noise. While some previous methods have been 
introduced for image enhancement, particularly in the field of medical 
imaging~\cite{zhang2023spectral}, they do not address the issue of noisy images. In fact, these approaches tend to struggle when dealing with images that have low peak signal-to-noise ratio (PSNR) as the noise is often amplified alongside the signal. For example, there exist some other image denoising methods~\cite{yang2018low}, using edge detection to enhance the structure feature of the original image while using different filters to remove the random noise. There is also another line.

\section{Method}

\begin{figure}[t]
    \centering
    \includegraphics[width=\columnwidth, clip=true, trim=0 5 0 2]{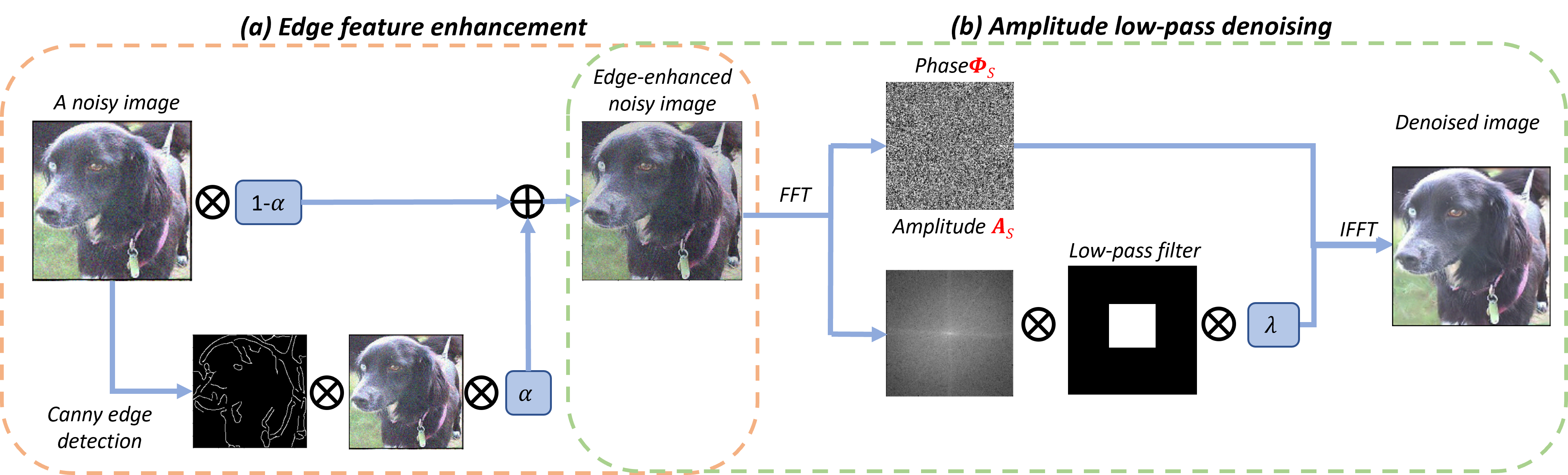}
    \caption{Method Overview.
}
    \label{fig:fig1}
\end{figure}

Deep learning has emerged as a promising approach to enhance the performance of denoising. used in image processing and computer vision, 
such as image filtering, edge detection, and feature extraction has been 
proposed. These techniques can be used in image denoising, for example, to 
identify edges in an image and preserve their structure while removing noise. 
And we learnt to use different types of filters for different kinds of noises. 
There have been many research papers published on image denoising, and 
various approaches have been proposed to solve this problem. Some 
popular methods include wavelet-based denoising, non-local means 
denoising, and sparse representation-based denoising.  
Among all the methods, edge enhancing method is a challenging yet crucial 
approach. Many techniques proposed in previous studies, such as histogram 
equalization, are sensitive to noise and can unintentionally enhance both the 
image structures and the noise. While some previous methods have been 
introduced for image enhancement, particularly in the field of medical 
imaging, they do not address the issue of noisy images. In fact, these 
approaches tend to struggle when dealing with images that have low peak 
signal-to-noise ratio (PSNR) as the noise is often amplified alongside the 
signal. 
For example, there exist some other image denoising methods, using 
edge detection to enhance the structure feature of the original image while 
using different filters to remove the random noise. There is also another line 2 of work, revealed that image structure information is mainly contained 
in the phase spectrum while the texture information is in the amplitude 
spectrum. Following these two directions, we propose a novel approach to 
do the image denoising. On one hand, we use the edge detection approach 
to discover and enhance the image structure information, on the other hand, 
we do the Fourier transform to get the amplitude and phase spectrum of the 
original images and apply different image denoising filters only on the 
amplitude spectrum. And then, we use inverse Fourier transform to 
reconstruct the denoised images. 
In addition to academic research, there are also consumer devices that 
incorporate image denoising techniques, such as smartphones with 
advanced camera systems that use software algorithms to remove noise 
from images captured in low-light conditions.

The overview diagram of our proposed method is presented in Fig. 1. 
Our method contains two major blocks. The first part utilizes a canny edge 
detection to detect the original edge features in the image and added the 
edge features back to enhance the semantic meaning of the original image 
in the box (a) of Fig.~\ref{fig:fig1}. Second part is the edge enhancement the amplitude 
spectrum denoising module, which is presented in  Fig.~\ref{fig:fig1}(b).

The pseudo code of the proposed framework is presented as below.

\noindent$===========================================================$
\begin{description}
\item[Input:] noisy image \(x\), weighting factor \(\alpha\) and \(\lambda\).
\item[Step 1:] Edge detection \(x_{\text{edge}} = \text{canny}(x)\)
\item[Step 2:] Edge enhancement \(x_{\text{enhance}} = x \cdot (1 - \alpha) + x \cdot \alpha \cdot x_{\text{edge}}\)
\item[Step 3:] FFT \((\phi, A) = \text{FFT}(x_{\text{enhance}})\)
\item[Step 4:] Amplitude low pass filtering \(A' = A \cdot M \cdot \lambda\)
\item[Step 5:] Image reconstruction \(x_{\text{denoised}} = \text{IFFT}(\phi, A')\)
\textbf{Output:} denoised image \(x_{\text{denoised}}\)
\end{description}
$===========================================================$

\section{Experimental Results}

The evaluation metrics we used to assess performance were RMSE (root mean squared error) and test \( R^2 \) which is defined as \( R^2 = 1 - \frac{(SSE/SST)} = 1 - \left(\frac{\text{Nest}(RMSE)^2}{\sum_{i=1}^{Nest}(y_i - \bar{y})^2}\right) \) to compare predictions to true scores. We created predictions for the test set and then calculate our metrics with respect to the predicted and observed values (i.e., the scores by the humans who graded the essay). We used test \( R^2 \) because it has a maximum of 1, providing an easy metric to understand model performance (higher values are better). For each dataset, we made sure that the ranges of scores were adapted to be the same (i.e., out of 100 with the same mean). For the essay models, we mean-adjusted because different exams might have different expected means, but the results should be centered around a known average. In practice, exam graders know what the overall average performance for a given exam would be. For the ELL dataset, because the output is 6 category scores, we averaged these to get a single essay score (to align with the FCE test set and our end goal) as shown in Tabl~\ref{table:essay_models}.

\begin{table}[ht]
\centering
\caption{Denoising Models}
\begin{tabular}{lcccrr}
\hline
\textbf{Model} & \textbf{Additional Dropout?} & \textbf{Nonlinearity} & \textbf{Data Used} & \textbf{RMSE} & \textbf{Test \(R^2\)} \\
\hline
Baseline & No & None & ICNALE & 11.1732 & 0.3255 \\
Basic & No & None & ELL & 10.8779 & 0.3607 \\
Basic & Yes & ReLU & ICNALE & 10.3576 & 0.4204 \\
Basic & Yes & ReLU & ELL & 10.9572 & 0.3513 \\
Basic & Yes & Convolution & ICNALE & 12.0160 & 0.2199 \\
Iterative & No & None & ELL, ICNALE & 10.7399 & 0.3768 \\
Iterative & Yes & ReLU & ETS, ELL & 10.8381 & 0.3654 \\
Hierarchical & Yes & None & ELL, ICNALE & 13.1572 & 0.0647 \\
Multi-task & Yes & None & ELL, ICNALE & 10.6521 & 0.3870 \\
\textbf{Ensemble} & \textbf{Yes} & \textbf{None} & \textbf{ELL, ICNALE} & \textbf{10.3125} & \textbf{0.4254} \\
\hline
\end{tabular}
\label{table:essay_models}
\end{table}

\section{Conclusions}

The field of image denoising continues to face increasing complexity 
and demands, underscoring the need for ongoing research. In this paper, we 
introduce our edge feature enhancement method and provide a 
comprehensive analysis of its methodology, including a comparison of its 
performance against other baseline methods. Going forward, we aim to 
investigate the effectiveness of our approach on a broader range of noisy 
image types and explore the potential for incorporating complementary 
techniques to enhance its capabilities further. Additionally, we plan to 
explore the possibility of combining different image denoising methods to 
determine if we can achieve even better results.

%
%
%
%
\newpage
\bibliographystyle{splncs04}
\bibliography{refs}
\end{document}